\documentclass[twocolumn]{revtex4-1}

\usepackage{amsmath}
\usepackage{setspace}
\usepackage{graphicx} 
\usepackage{capt-of} 
\usepackage{amstext}
\usepackage[ngerman,english]{babel}
\usepackage{xcolor}
\usepackage{physics}
\usepackage{booktabs}
\usepackage{soul}
\soulregister\cite7
\soulregister\ref7
\usepackage{hyperref} 
\hypersetup{colorlinks=false, citebordercolor=green}

\begin{document}
\title{Interaction induced single impurity tunneling in a binary mixture of trapped ultracold bosons}

\author{Kevin Keiler $^1$}
\author{Peter Schmelcher $^{1,2}$}
\affiliation{$^1$Center for Optical Quantum Technologies, University of Hamburg, Department of Physics, Luruper Chaussee 149, 22761 Hamburg, Germany}
\affiliation{$^2$The Hamburg Centre for Ultrafast Imaging, University of Hamburg, Luruper Chaussee 149, 22761 Hamburg, Germany}


\begin{abstract}
	We investigate the tunneling dynamics of an ultracold bosonic impurity species which interacts repulsively with a second, larger Bose gas. Both species are held in a finite-sized quasi-one-dimensional box potential. In addition, the impurity bosons experience a periodic potential generated by an optical lattice. We initially prepare our binary mixture in its ground state, such that the impurities and Bose gas are phase separated and the impurities localize pairwise in adjacent sites of the periodic potential, by tuning the interaction strengths and the lattice depth correspondingly. The dynamics is initiated by suddenly lowering the repulsive interspecies interaction strength, thereby entering a different regime in the crossover diagram. For specific post-quench interspecies interaction strengths we find that a single impurity tunnels first to the neighbouring empty site and  depending on the quench strength can further tunnel to the next neighbouring site. Interestingly, this effect is highly sensitive to the presence of the Bose gas and does not occur when the Bose gas does not interact with the impurity species throughout the dynamics. Moreover, we find that the tunneling process is accompanied by strong entanglement between the Bose gas and the impurity species as well as correlations among the impurities. 
\end{abstract}
\maketitle

\section{Introduction}
\label{sec:Introduction}
Ultracold atoms have proven to represent a very flexible platform for exploring many-body quantum effects. They allow for varying the trapping potential in numerous ways, ranging from the manipulation of the dimensionality to 'painting' arbitrary external confinements \cite{boshier}. Additionally, Feshbach and confinement induced resonances \cite{confine,feshbach} enable the tuning of the interaction strength  between the atoms. In this spirit ultracold atoms allow to realize even rather complex many-body systems. In particular, one-dimensional (1D) systems are under intense investigation since they exhibit intriguing phenomena. Due to their inverse scaling of the effective interaction strength to the density \cite{1d_eff_corr_1,1d_eff_corr_2,pinning} they allow for entering the strong interaction regime, while remaining dilute \cite{1d_1,1d_2}. Consequently, 1D ultracold systems provide us with the possibility to enter regimes where effective theories are no longer valid.\par 
Apart from controlling single component fermionic or bosonic ensembles in a systematic manner it is possible to extend this control to mixtures of ultracold atoms \cite{spec_el,wiemanRB} such as Bose-Bose, Fermi-Fermi and Bose-Fermi mixtures. The fact that we add to the already present intra-species interactions an interspecies interaction leads to a significant extension of the already very rich phenomenology of the case of a single species, ranging from pair-tunneling \cite{pflanzer1,pflanzer2} to phase separation processes \cite{phase_sep1,ofir,phase_sep2,phase_sep3} and composite fermionization \cite{cf,pyzh,zollner}. Here, especially particle-imbalanced mixtures are currently of immediate interest. Such a setup can consist of an impurity species of a few particles which are immersed in a majority species of many particles. Extensive theoretical \cite{blume_imp,cuc_imp,massignan_review,grusdt,volosniev,zinner_pol,garcia,lampo1,lampo2,knoerzer,lia2018,simortho} as well as experimental \cite{corn_pol_sim,trans_imp,naegerl_bloch,catani,fukuhura} studies have been performed for this case. A single impurity immersed in a majority species leads to the notion of a polaron, which plays a central role in our understanding of quantum matter. Considering several impurities interacting with a majority species, one finds that the latter mediates an effective attractive interaction between the impurities \cite{zwerger_casimir,fleischhauer_ind,zinner_ind,jie_chen,jie2,simosfermi,bipol}, leading to a clustering of the impurities assuming they are bosons  \cite{jaksch_pol,jaksch_clus,jaksch_trans,keiler1}. As it turns out, it is even possible to engineer the localization properties of the impurities by including a repulsive intra-species interaction between them \cite{keiler2}. \par
In the present work, we investigate a binary Bose-Bose mixture consisting of an ultracold bosonic impurity species which interacts repulsively with a second, larger majority species. Both species are held in a finite-sized quasi-one-dimensional box potential. In addition, the impurity bosons experience a periodic potential generated by an optical lattice. Exploiting the aforementioned possibility to engineer the localization properties of the impurities, we prepare an initial wave function which constitutes a phase separation of the impurity species and the Bose gas, while the impurities accumulate pairwise in adjacent wells of the lattice potential. Upon lowering the interspecies interaction strength we explore the dynamical response of the system. The corresponding quench is performed across crossover boundaries of regimes which constitute different distributions of the impurities in the lattice depending on the interspecies interaction strength. Building upon this, we aim at a dynamical particle transfer, which takes place in a systematic manner due to the presence of the majority species. We find that a sufficiently strong quench leads to the tunneling of a single impurity through the lattice out of the initial four-impurities cluster. Interestingly, this process is only possible for a finite post-quench interspecies interaction strength and does not appear when the Bose gas is transparent to the impurity species in the course of time. We observe that the tunneling process of the impurity species is accompanied by a strong entanglement between the subsystems as well as strong correlations among the impurities. Increasing the number of particles in the majority species, we observe that the tunneling of the impurity species persists, while exhibiting a slightly lower tunneling amplitude.\par
Our work is structured as follows: In section \ref{sec:setup_methodology} we present the system under investigation along with the numerical method employed for simulating the correlated many-body dynamics. Section \ref{sec:Quench Protocol} provides a discussion of the quench protocol as well as the initial ground state. Section \ref{sec:Tunneling Dynamics} is dedicated to a thorough analysis of the dynamical response as a function of the interspecies interaction strength and the number of majority species particles. We conclude in section \ref{sec:Conclusion} with a summary of our findings, present possible applications and discuss directions for future studies.

\section{Setup and computational approach}
\label{sec:setup_methodology}

\subsection{Computational Approach}
Our numerical simulations are performed using the \textit{ab-initio} Multi-Layer Multi-Configuration Time-Dependent Hartree method for bosonic (fermionic) Mixtures (ML-MCTDHX) \cite{mlb1,mlb2,mlx}, which is able to take all correlations into account \cite{bb2018,ff2018,ff2019,bf2018,lode1,lode2}.
As a first step, the total many-body wave function $\ket{\Psi_{\textrm{MB}}(t)}$ is expanded in $M$ species functions $\ket{\Psi^{\sigma}(t)}$ of species $\sigma$ and written as a Schmidt decomposition \cite{schmidt_dec} 
\begin{equation}
	\ket{\Psi_{\textrm{MB}}(t)} = \sum_{i=1}^{M} \sqrt{\lambda_{i}(t)} \ket{\Psi_{i}^A(t)}\otimes \ket{\Psi_{i}^B(t)}.
	\label{eq:schmidt}
\end{equation}
Here, the Schmidt coefficients $\sqrt{\lambda_{i}}$, in decreasing order, provide information about the degree of population of the $i-$th species function and thereby about the degree of entanglement between the impurities and the majority species. In case that $\lambda_1=1$ the species $A$ and $B$ are not entangled and the system can be described with a species mean-field ansatz ($M=1$).

Furthermore, the species wave functions $\ket{\Psi^{\sigma}(t)}$ describing an ensemble of $N_\sigma$ bosons are expanded in a set of permanents
\begin{equation}
	\ket{\Psi_{i}^\sigma(t)} = \sum_{\vec{n}^\sigma|N_\sigma} C_{\sigma\vec{n}}(t)
	|\vec{n}^\sigma;t\rangle,
	\label{eq:ml_ns}
\end{equation}
where the vector $\vec{n}^\sigma=(n^{\sigma}_1,n^{\sigma}_2,...)$ denotes the occupations of the time-dependent single-particle functions of the species $\sigma$. The notation $\vec{n}^\sigma|N_\sigma$ indicates that for each $|\vec{n}^\sigma;t\rangle$ we require the condition $\sum_{i}n^{\sigma}_i=N_\sigma$.
The time propagation of the many-body wave function is achieved by employing the Dirac-Frenkel variation principle $ \bra{\delta\Psi_\textrm{MB}} (\textrm{i}\partial_t - \mathcal{H} )\ket{\Psi_\textrm{MB}} $ \cite{var1,var2,var3} with the variation $\delta\Psi_\textrm{MB}$.
Within ML-MCTDHX one has access to the complete many-body wave function which allows us consequently to derive all relevant characteristics of the underlying system. In particular, this means that we are able to characterize the system by projecting onto number states w.r.t. an appropriate single-particle basis \cite{ns_analysis1,ns_analysis2}. Besides investigating the quantum dynamics it allows us to calculate the ground (or excited) states by using either imaginary time propagation or improved relaxation \cite{meyer_improved}, thereby being able to uncover also possible degeneracies of the many-body states. In standard approaches for solving the time-dependent Schr{\"o}dinger equation, one typically constructs the wave function as a superposition of time-independent Fock states with time-dependent coefficients. Instead, it is important to note that the ML-MCTDHX approach considers a co-moving time-dependent basis on different layers, meaning that in addition to time-dependent coefficients the single particle functions spanning the number states are also time-dependent. This leads to a significantly smaller number of basis states and configurations that are needed to obtain an accurate description and thus reduces the computation time \cite{pruning}.

\subsection{Setup}
Our system consists of a mixture of two bosonic species. The minority species A (impurities) is trapped in a one-dimensional lattice with hard wall boundary conditions. It is immersed in a majority species B obeying the same boundary conditions but without the lattice potential. This setup lies within reach of current experimental techniques. Various trapping potentials for the atoms can be achieved, including in particular one-dimensional ring geometries \cite{boshier} and box potentials \cite{box_pot}. The optical lattice potential for the impurities does not affect the Bose gas, which is achievable by choosing the corresponding laser wavelengths and atomic species \cite{spec_sel_lat}. Thereby, we create a two-component system with each species being trapped individually. Furthermore, we introduce a coupling Hamiltonian $\hat{H}_{AB}$ between the two species. Both subsystems are confined to a longitudinal direction, accounting for the one-dimensional character, and excitations in the corresponding transversal direction are energetically suppressed and can therefore be neglected. This results in a Hamiltonian of the form $\hat{H}=\hat{H}_A+\hat{H}_B+\hat{H}_{AB}$.
The Hamiltonian of the A species reads
\begin{equation}
	\begin{split}
		\hat{H}_A&=\int_{-L/2}^{L/2} \text{dx} \, \hat{\Psi}_{A}^{\dagger}(\text{x}) \Big [ -\frac{\hbar^{2}}{2 m_A} \frac{\text{d}^{2}}{\text{dx}^{2}}+ V_0 \sin^{2}\Big(\frac{\pi k \text{x}}{L}\Big) \\
		&+ g_{AA} \; \hat{\Psi}_{A}^{\dagger}(\text{x}) \hat{\Psi}_{A}(\text{x}) \Big ] \hat{\Psi}_{A}(\text{x}),
	\end{split}
\end{equation}
where $\hat{\Psi}_{A}^{\dagger}$ is the field operator of the lattice A bosons, $m_A$ their mass, $V_0$ the lattice depth, $g_{AA}$ the intraspecies interaction strength of the two-body contact interaction among the A atoms, $k$ the number of wells in the lattice and $L$ is the length of the system, while $x\in[-L/2,L/2]$.
The B species is described by the Hamiltonian
\begin{equation}
	\begin{split}
		\hat{H}_B&=\int_{-L/2}^{L/2} \text{dx} \; \hat{\Psi}_{B}^{\dagger}(\text{x}) \Big [ -\frac{\hbar^{2}}{2 m_B} \frac{\text{d}^{2}}{\text{dx}^{2}} \\
		&+ g_{BB} \; \hat{\Psi}_{B}^{\dagger}(\text{x})\hat{\Psi}_{B}(\text{x}) \Big ] \hat{\Psi}_{B}(\text{x}),
	\end{split}
\end{equation}
where $\hat{\Psi}_{B}^{\dagger}$ is the field operator of the B species, $g_{BB}>0$ is the interaction strength of the two-body contact interaction among the B atoms and $m_B$ is the corresponding mass.  Moreover, we assume equal masses for the species $m_A=m_B$. Experimentally this can be achieved by preparing e.g. $^{87}{Rb}$ atoms in two different hyperfine states, i.e. $|F= 2,m_F=-2\rangle$ represents the impurity species and  $|F= 1,m_F=-1\rangle$ represents the Bose gas, thereby obtaining a  two-species  bosonic  mixture.  Using the  so-called  'tune-out' wavelength of the $|F= 1,m_F=-1\rangle$ state it is possible to create species-dependent potentials \cite{spec_sel_lat,mbl_hyperfine}, such that the optical lattice potential for the impurities does not affect the Bose gas. Consequently, the Bose gas experiences a vanishing light shift, while the light shift for the impurity species can be used to trap this species in an optical lattice potential. The interaction between the species A and B is given by
\begin{equation}
	\hat{H}_{AB}= g_{AB} \int_{-L/2}^{L/2} \text{dx} \; \hat{\Psi}_{A}^{\dagger}(\text{x}) \hat{\Psi}_{A}(\text{x}) \hat{\Psi}_{B}^{\dagger}(\text{x}) \hat{\Psi}_{B}(\text{x}),
\end{equation}
where $g_{AB}$ is the interspecies interaction strength. The interaction strengths $g_{\alpha}$ ($\alpha\in\{A,B,AB\}$) can be expressed in terms of the three dimensional s-wave scattering lengths $a^{3D}_{\alpha}$. By assuming the above-mentioned strong transversal confinement with the same trapping frequencies $\omega^{\sigma}_{\perp}=\omega_{\perp}$ for both species $\sigma \in \{A,B\}$ it is possible to integrate out frozen degrees of freedom, leading to a quasi one-dimensional model with $g_{\alpha}=2\hbar\omega_{\perp}a^{3D}_{\alpha}$.\par
Throughout this work we consider a $k=5$ well lattice and focus on small particle numbers with $N_A=4$ impurities, leading thus to a fractional filling in the lattice, and $N_B\in\{10,30\}$ majority atoms. The interaction among the majority atoms is set to a value where the quantum depletion of the Bose gas is negligible in case of an absent interspecies coupling, i.e. $g_{BB}/E_R \lambda=6.8 \times 10^{-3}$, with $E_R=(2\pi\hbar)^{2}/2m_A \lambda^{2}$ being the recoil energy and $ \lambda=2L/k$ the optical lattice wavelength.
\begin{figure}[t]
	\includegraphics[width=0.9\columnwidth]{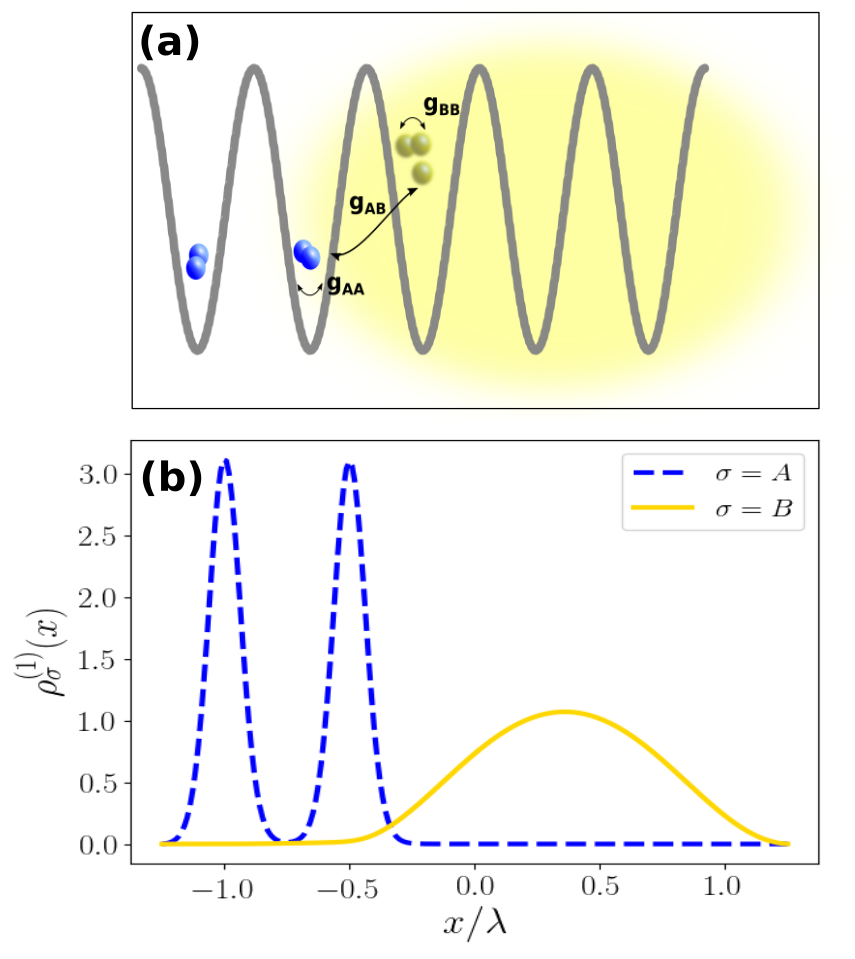}
	\caption{(a) Sketch of the initial state of the two-component mixture. The impurities [blue (dark shading) dots] interact repulsively via an intraspecies contact interaction of strength $g_{AA}$ and via an interspecies contact interaction of strength $g_{AB}$ with the atoms of the Bose gas. The latter atoms in turn weakly interact via an intraspecies contact interaction of strength $g_{BB}$.  Due to the interplay between the interspecies and the intraspecies interaction among the impurities the latter accumulate pairwise in adjacent sites, minimizing the overlap with the majority species. (b) One-body density of the many-body ground state of the $\sigma$ species $\rho^{(1)}_\sigma(x,t)$ for $N_A=4$, $N_B=10$, a lattice depth of $V_0/E_R=13$, intraspecies interaction strengths of $g_{AA}/E_R\lambda=0.067$ and $g_{BB}/E_R\lambda=6.8 \times 10^{-3}$ and an interspecies interaction strength of $g_{AB}/E_R\lambda=0.142$. $x$ is given in units of $\lambda$.}
	\label{fig:initial_state}
\end{figure}
In the following, we present the quench protocol which induces the tunneling dynamics. We prepare our system in its ground state with $g_{AA}/E_R \lambda=0.067$, $g_{AB}/E_R \lambda=0.142$ and $V_0/E_R=13$. This choice of parameters ensures that we arrive approximately at the following many-body (MB) wave function $|\Psi_{\text{MB}}\rangle\approx|2,2,0,0,0\rangle_W\otimes|\Psi_B\rangle$, where the impurities accumulate pairwise in adjacent sites (cf. Figure \ref{fig:initial_state}). The number state $|2,2,0,0,0\rangle_W$ is spanned by Wannier states of the lowest band for the impurity species, whereas $|\Psi_B\rangle$ denotes the species wave function of the majority species. Hence, initially the subsystems are not entangled, thereby forming a single product state. We further note that the two species strongly avoid overlap in this ground state configuration (see \cite{keiler2}), reminiscent of the phase separation of two Bose gases \cite{phase_sep1}. This can be seen in the one-body density of the ground state $|\Psi_{\text{MB}}\rangle$ of the species $\sigma$ [cf. Figure \ref{fig:initial_state} (b)], which is defined as
\begin{equation}
	\rho^{(1)}_\sigma(x)=\langle\Psi_\text{MB}| \hat{\Psi}_{\sigma}^{\dagger}(x)\hat{\Psi}_{\sigma}(x)|\Psi_\text{MB} \rangle.
\end{equation}
The quench is performed by lowering the interspecies interaction strength. Varying the post-quench interspecies interaction strength $g_{AB}$, we explore the dynamical response of the binary mixture, focusing in particular on the tunneling behaviour of the impurity species.

\section{The Quench Protocol and its underlying physics}
\label{sec:Quench Protocol}
\subsection{Crossover Diagram}
Before discussing the dynamical response of the binary mixture upon quenching the interspecies interaction strength $g_{AB}$, we shall motivate the employed quench protocol. For this purpose, we determine the ground state of the system for different lattice depths $V_0$ and $g_{AB}$. We note that the underlying single particle functions of the ground state are optimized w.r.t. the respective many-body problem and can in general be of arbitrary structure.
\begin{table*}[t]
	\captionof{table}{Degenerate subspaces of the ground state referring to the regimes in Figure \ref{fig:crossover} (d) for $N_B=10$ and $V_0/E_R=13$. }
	\label{table_states}
	\begin{tabular}{ccc}
		\hline\hline
		Regime: $I$  & $II$ &  $III$ \\\\ \hline
		$|1,1,0,1,1\rangle_W\otimes|\Psi^1_B\rangle$  $\;$ & $|2,1,0,0,1\rangle_W\otimes|\Psi^{2}_B\rangle$ $\;$ & $|2,2,0,0,0\rangle_W\otimes|\Psi^{3}_B\rangle $ \\\\
		$\;$&  $|1,0,0,2,1\rangle_W\otimes|\bar{\Psi}^2_B\rangle$  $\;$& $|0,0,0,2,2\rangle_W\otimes|\bar{\Psi}^3_B\rangle$ \\
		\hline\hline
	\end{tabular}
\end{table*} 
\begin{figure}[t]
	\includegraphics[width=\columnwidth]{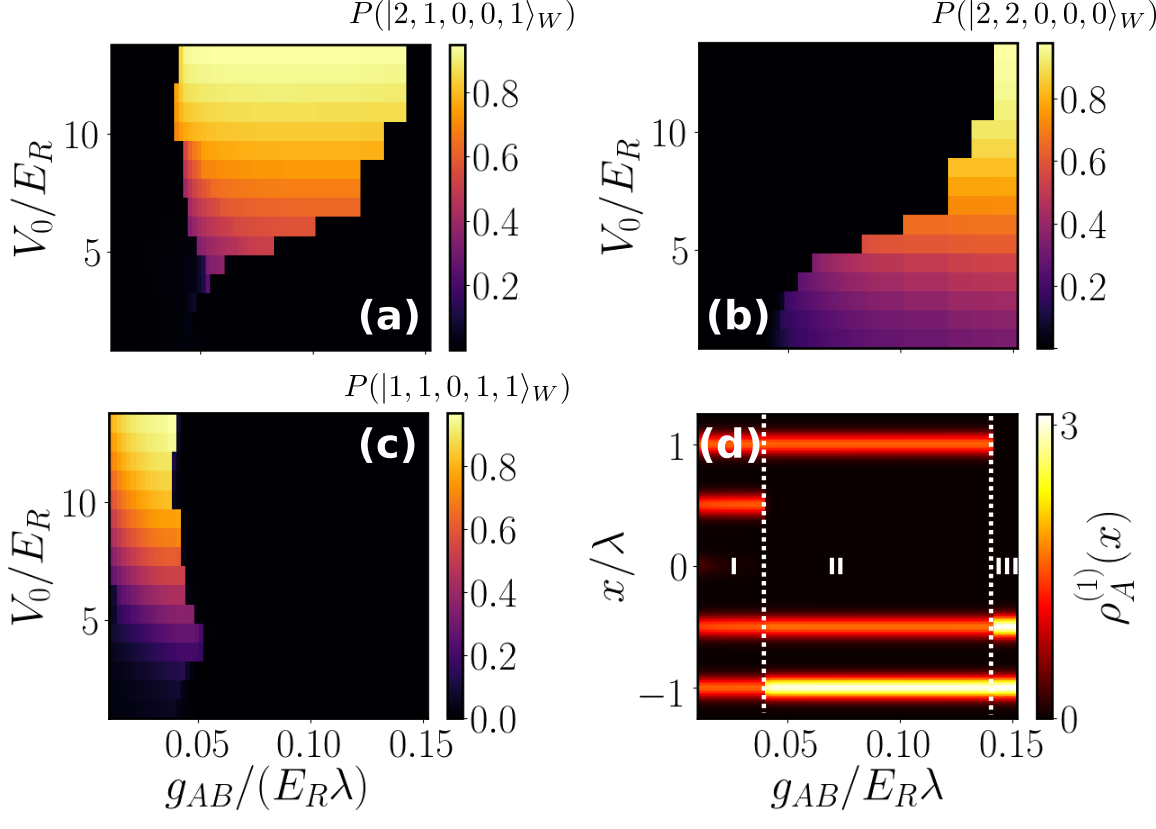}
	\caption{Probability of finding the impurity species (a) forming one two-impurity cluster next to a single impurity and one impurity residing along the opposite wall, (b) localized pairwise in adjacent wells and (c) being localized separately in the outer wells as a function of the lattice depth $V_0$ and the interspecies interaction strength $g_{AB}$ for $g_{AA}/E_R \lambda=0.067$. The number states in (a) and (b) are chosen as an example and therefore the corresponding parity symmetric configurations show the same behaviour. In (d) we show the one-body density of the species A as a function of the interspecies interaction strength for a fixed lattice depth of $V_0/E_R=13$. In region I the impurities localize separately in the outer wells. In region II they form one two-impurity cluster next to a single impurity and one impurity residing along the opposite wall and in region III they are localized pairwise in adjacent wells (see Table \ref{table_states}). The particle number of the respective species is chosen as $N_A=4$ and $N_B=30$. $x$ is given in units of $\lambda$, $V_0$ in units of $E_R$ and $g_{AB}$ in units of $E_R\lambda$.}
	\label{fig:crossover}
\end{figure}
We extract information from the complete many-body wave function by projecting onto number states $|\vec{n}^{A}\rangle\otimes|\vec{n}^{B}\rangle$ and determine the probability of being in the number state $|\vec{n}^{A}\rangle$ for the impurity species A, irrespective of the number state configurations of the B species, namely
\begin{equation}
	P(|\vec{n}^{A}\rangle)= \sum_{i} |\langle \vec{n}^{B}_i |\otimes\langle \vec{n}^{A}|\Psi_{\text{MB}}\rangle|^{2},
	\label{eq:prob_A}
\end{equation}
where $\{|\vec{n}^{B}_i\rangle\}$ could be any number state basis set of the Bose gas with fixed particle number and $|\Psi_{\text{MB}}\rangle$ is the total many-body ground state wave function. 
In order to associate the impurity state $|\vec{n}^{A}\rangle=|n^A_1,n^A_2,n^A_3,n^A_4,n^A_5\rangle$ with a spatial distribution we construct the number states with a generalized Wannier basis (cf. subscript W) of the lowest band \cite{kivelson1,kivelson2,footnote1}.\par 
In Figure \ref{fig:crossover} we show the probability $P(|\vec{n}^{A}\rangle)$ of finding the impurity A species atoms in the respective number state $|\vec{n}^{A}\rangle$ as a function of  $V_0$ and $g_{AB}$. Predominantly the states $|1,1,0,1,1\rangle_W$, $|2,1,0,0,1\rangle_W$  and $|2,2,0,0,0\rangle_W$  are occupied. Focusing on large lattice depths the impurity species is well described by a single number state of the aforementioned kind depending on the interspecies interaction strength. For small $g_{AB}$ and large lattice depths we observe that the impurities are localized separately in the outer wells [cf. Figure \ref{fig:crossover} (c)]. Increasing $g_{AB}$ we find a transition to a state exhibiting one two-impurity cluster next to a single impurity and one impurity residing along the opposite wall [cf. Figure \ref{fig:crossover} (a)]. For even larger $g_{AB}$, the impurities accumulate pairwise in adjacent sites [cf. Figure \ref{fig:crossover} (b)]. This behaviour is also reflected in the one-body density $\rho^{(1)}_{A}(x)$ of the impurity species, shown in Figure \ref{fig:crossover} (d). We note here that the many-body ground states in Figure \ref{fig:crossover} (a) and (b) for large lattice depths are degenerate with an energetically equivalent counterpart of the type $|1,0,0,1,2\rangle_W\otimes|\bar{\Psi}^2_B\rangle$ or $|0,0,0,2,2\rangle_W\otimes|\bar{\Psi}^3_B\rangle$, where $|\bar{\Psi}^i_B\rangle$ and $|\Psi^i_B\rangle$ denote the respective species wave function of the majority species B (the bar denotes the species wave function B for the degenerate counterpart). For large lattice depths we can now classify the ground states for varying interspecies interaction strengths into three regimes as depicted in Table \ref{table_states}.\par
Next, we prepare the binary mixture in the ground state many-body wave function $|\Psi_0\rangle\approx|2,2,0,0,0\rangle_W\otimes|\Psi^{3}_B\rangle$ by choosing a lattice depth of $V_0/E_R=13$ and an interspecies interaction strength of $g_{AB}/E_R\lambda=0.142$. Lowering now instantaneously the interspecies interaction strength and thereby crossing the different regimes in the crossover diagram in Figure \ref{fig:crossover}, the corresponding ground states would be given by more delocalized configurations (cf. Table \ref{table_states}). In this spirit, one might expect that the former two-impurity cluster will split up in the course of the dynamics, leading to a tunnelling of the impurities in the lattice. Of course, this simple picture will not give insight into the explicit complex dynamical response of our system, but rather serves as a starting point as well as a motivation for the chosen quench protocol. An extensive study of the actual dynamics observed in the two-component system will be the subject of the following sections.
\subsection{Single Impurity Dynamics}
Instead of diving directly into the results for the case of $N_A=4$ impurities in the subsystem A, we will first briefly discuss the case of a single impurity, i.e. $N_A=1$, coupled to a majority species of $N_B=10$ particles. Assuming the same parameters as discussed above and setting the lattice depth to $V_0/E_R=13$ and the interspecies interaction strength to $g_{AB}/E_R\lambda=0.14$, the impurity will accumulate to one of the most outer lattice sites, similar to the case of $N_A=4$ impurities. Quenching now the interspecies interaction strength to a lower value, we find two distinct regimes for the dynamical response of the impurity, identified by the temporal evolution of the one-body density for both species upon quenching $g_{AB}$.
\begin{figure}[t]
	\includegraphics[width=\columnwidth]{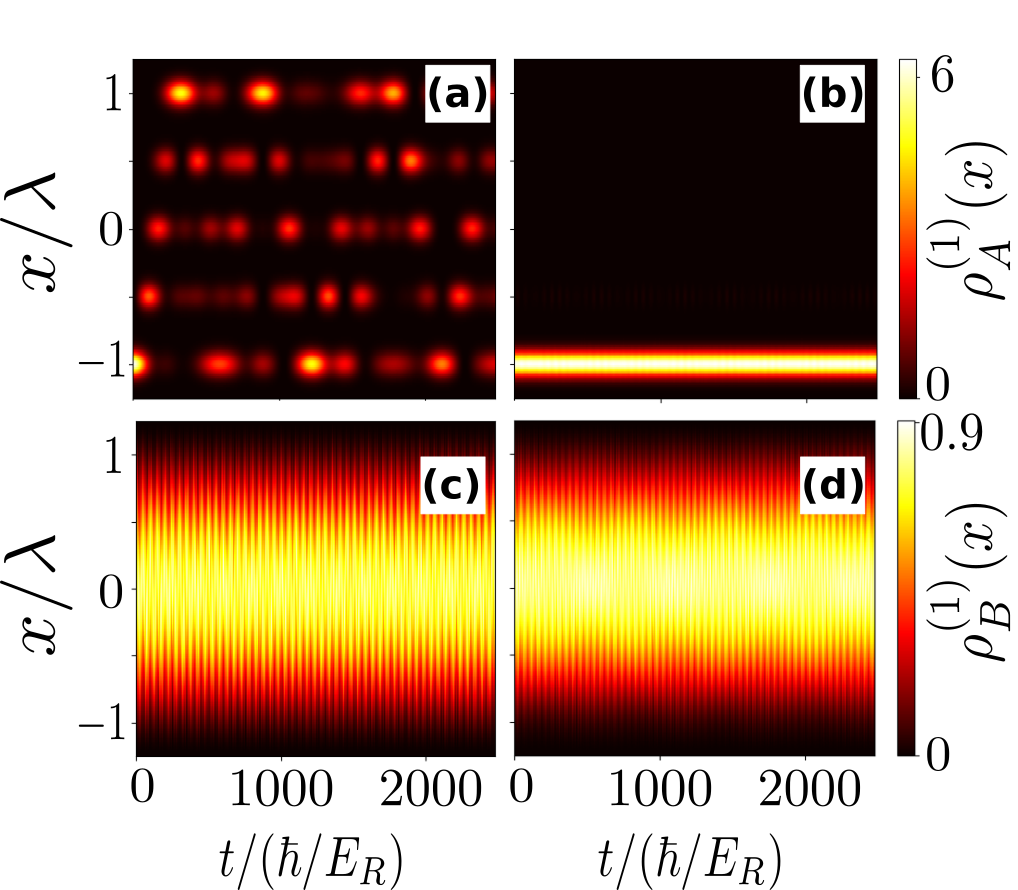}
	\caption{Temporal evolution of the one-body density of the A species $\rho^{(1)}_A(x,t)$ (upper row) and B species $\rho^{(1)}_B(x,t)$ (lower row) for a single impurity coupled to the majority species B for a lattice depth of $V_0/E_R=13$ and a post-quench interspecies interaction strength of (a), (c) $g_{AB}/E_R\lambda=0$ and (b), (d) $g_{AB}/E_R\lambda=0.032$. $x$ is given in units of $\lambda$ and $t$ in units of $\hbar/E_R$.}
	\label{fig:single_impurity}
\end{figure}
We observe that initially the impurity is localized in the rightmost well, whereas the majority species distributes such that it avoids the overlap with the impurity. For a post-quench interspecies interaction strength of $g_{AB}/E_R\lambda=0$, we observe a tunneling of the impurity through the lattice potential structure to the leftmost well [cf. Figure \ref{fig:single_impurity} (a)]. This is followed by a rather complex tunneling between the various lattice sites. Throughout the dynamics the majority species exhibits high-frequency oscillations of the one-body density  [cf. Figure \ref{fig:single_impurity} (c)].
For any post-quench interspecies interaction strength not close to zero the impurity remains localized in the initially populated well in the course of the dynamics [cf. Figure \ref{fig:single_impurity} (b)], whereas the B species exhibits again high-frequency oscillations on top of an overall breathing of the one-body density [cf. Figure \ref{fig:single_impurity} (d)]. Apparently, the finite repulsive coupling of the impurity to the majority species does not allow for a transfer of the impurity to the neighbouring lattice sites due to the latter species acting as an effective material barrier. Instead it is necessary to quench the system such that the majority species basically becomes transparent to the impurity in order achieve a tunneling of the impurity.
In this sense, the dynamical response of the system can be defined by two distinct regimes. Either the impurity performs a rather complex tunneling through the lattice potential, while exhibiting traces of an oscillation from one side to the other side of the system, for post-quench $g_{AB}=0$ or close to zero, or it remains localized in the initially populated well for weaker quench amplitudes. 
In the following discussion we will explicitly show that the case of four impurities initially accumulating pairwise in adjacent sites exhibits impurity tunneling for finite post-quench couplings to the majority species and strong enough quench amplitudes, whereas quenching to $g_{AB}/E_R\lambda=0$ results in a localization of the impurity species in the initially populated wells in the course of time.

\section{Correlated Tunneling Dynamics}
\label{sec:Tunneling Dynamics}
\subsection{Density Evolution and Correlation Analysis}
\label{sec:Tunneling DynamicsA}
\begin{figure*}[t]
	\includegraphics[width=1.9\columnwidth]{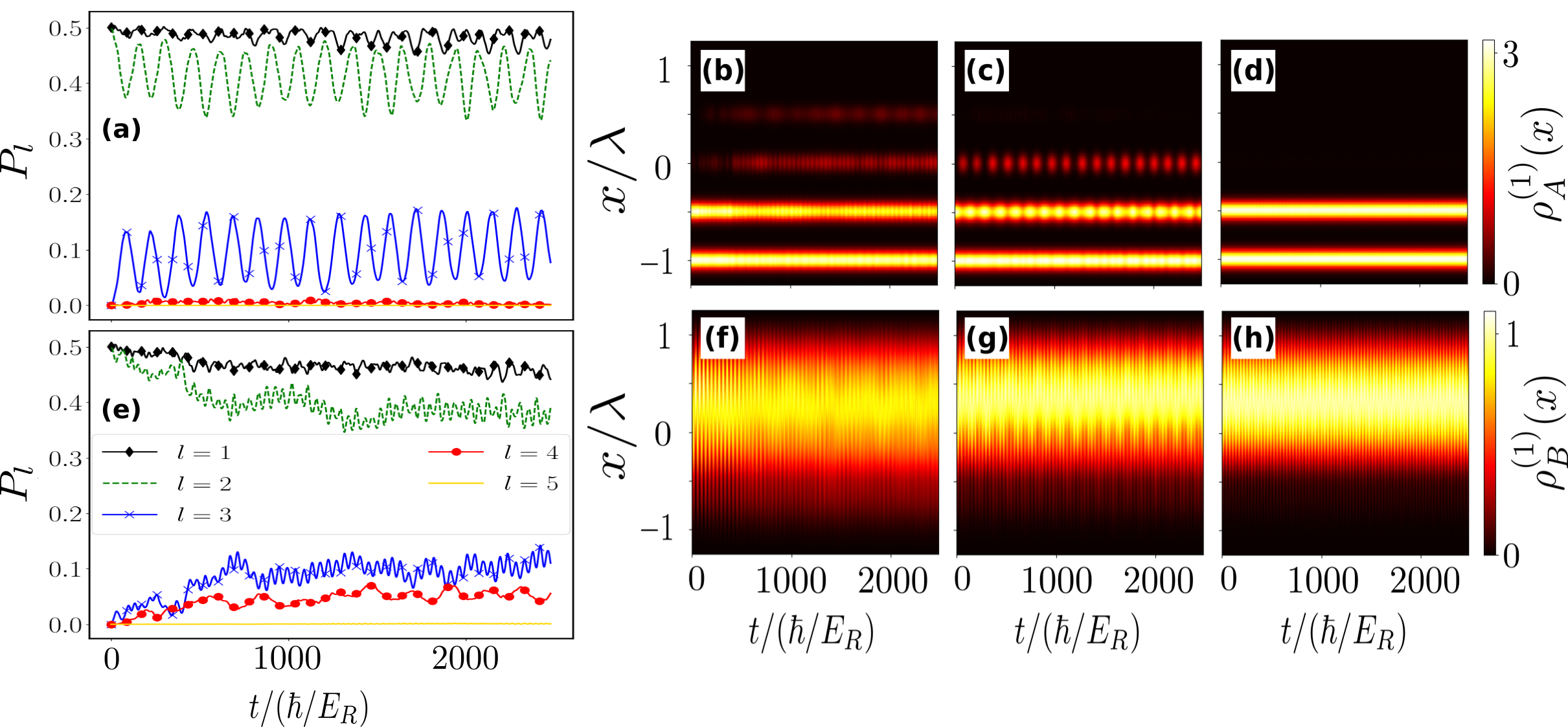}
	\caption{Temporal evolution of the probability $P_l$ of finding an impurity particle in the $l-$th Wannier state upon quenching to (a) $g_{AB}/E_R\lambda=0.051$ and (e) $g_{AB}/E_R\lambda=0.032$. Temporal evolution of the one-body density of the A species $\rho^{(1)}_A(x,t)$ (upper row) and B species $\rho^{(1)}_B(x,t)$ (lower row) for  a lattice depth of $V_0/E_R=13$ and a post-quench interspecies interaction strength of (b), (f) $g_{AB}/E_R\lambda=0.032$, (c), (g) $g_{AB}/E_R\lambda=0.051$ and (d), (h) $g_{AB}/E_R\lambda=0.065$. The particle number of the respective species is chosen as $N_A=4$ and $N_B=10$. $x$ is given in units of $\lambda$ and $t$ in units of $\hbar/E_R$.}
	\label{fig:densities}
\end{figure*}
We explore the dynamics of $N_A=4$ impurities coupled to a majority species of $N_B=10$ bosons. Motivated by the crossover diagram in the previous section [cf. Figure \ref{fig:crossover}] we choose a lattice depth of $V_0/E_R=13$ and an interspecies interaction strength of $g_{AB}/E_R\lambda=0.14$ in order to prepare the binary mixture in the ground state $|\Psi_0\rangle\approx|2,2,0,0,0\rangle_W\otimes|\Psi^{3}_B\rangle$. We perform a quench of the interspecies interaction strength to a lower value and determine the dynamical response of the system numerically using ML-MCTDHX which is capable of taking all correlations into account.
In the first step, we identify four different regimes of the dynamical response of our system, depending on the post-quench interspecies interaction strength $g_{AB}$, based on qualitatively different dynamical patterns of the one-body densities.
Figure \ref{fig:densities}, shows the temporal evolution of the one-body density of the A and B species for various post-quench interspecies interaction strengths. For weak quench amplitudes [cf. Figure \ref{fig:densities} (d), (h)], the impurities remain trapped in the initially occupied wells and appear not to respond dynamically to the quench, whereas the majority species performs high-frequency oscillations in the one-body density. The origin of these oscillations will be discussed in section \ref{sec:character}. This scenario resembles the case of a single impurity coupled to the majority, which has been discussed in the context of Figure \ref{fig:single_impurity}. However, quenching to even lower values of $g_{AB}$ initiates a tunneling process to the neighbouring unoccupied lattice site [cf. Figure \ref{fig:densities} (c)]. We observe a rather periodic oscillation of the one-body density $\rho^{(1)}_A(x,t)$ between the second and third lattice site. These oscillations are imprinted on the one-body density profile of the majority species, which tries to minimize the overlap to the minority species [cf. Figure \ref{fig:densities} (g)], on top of the high-frequency oscillations. We can extend this tunneling behaviour of the impurities by reducing the post-quench interspecies interaction strength even further. This leads to a tunneling of the impurity species into the third, as in the previous case, and in particular into the fourth well. Again, this will leave an imprint on the one-body density of the B species [cf. Figure \ref{fig:densities} (f)]. Remarkably, for a post-quench interspecies interaction strength of $g_{AB}=0$, such that the majority species is transparent to the impurities, the latter will remain localized in the initially populated wells (not shown here). In contrast to the case of a single impurity (see Figure \ref{fig:single_impurity}), the two-impurity cluster remains stable and does not deviate from its initial configuration in the course of time. Such stable composite objects exhibit rather large lifetimes and have already been observed experimentally in optical lattices \cite{rep_bound_pairs}. The majority species, however, does respond strongly to the quench which manifests itself in an initial dipole-like motion of the species, which stabilizes in time due to interference processes (not shown here).\par
In order to deepen our insight into the mechanisms underlying the dynamics in the binary mixture, we will now analyze the correlations which accompany the dynamics, in particular the entanglement between the impurity species and the majority species.
For this purpose we introduce the von Neumann entropy
\begin{equation}
	S_{AB}(t)=-\sum_{i} \lambda_i(t) \ln(\lambda_i(t))
	\label{eq:vonNeumann}
\end{equation}
as a measure for the entanglement between the subsystems A and B, where $\lambda_i$ are the Schmidt coefficients defined in equation \ref{eq:schmidt}. In the case of a single contributing product state in equation \ref{eq:schmidt}, the subsystems are disentangled and the von Neumann entropy is given by $S_{AB}=0$, whereas any deviation from this value indicates entanglement between the A and the B species. In the same manner we want to define a measure for the correlations which are present in each subsystem itself. Let us first remember the spectral decomposition of the one-body density of species $\sigma$ which reads
\begin{equation}
	\rho_\sigma^{(1)}(x,t) = \sum_j n_{\sigma j}(t) \Phi^{*}_{\sigma j}(x,t)\Phi_{\sigma j}(x,t),
	\label{eq:natural_populations}
\end{equation}
where $n_{\sigma j}(t)$ in decreasing order, obeying $ \sum_j n_{\sigma j}=1$, are the so-called natural populations and $\Phi_{j\sigma}(x,t)$ the corresponding natural orbitals. In this sense, the natural orbitals are the eigenstates, while the natural populations are the corresponding eigenvalues, which are determined by diagonalizing the one-body density matrix. Similar to the Schmidt coefficients the natural populations serve as a measure for the correlations in a subsystem. In this spirit, we define the fragmentation in the subsystem $\sigma$ as
\begin{equation}
	S_{\sigma}(t)=-\sum_{j} n_{\sigma j}(t) \ln(n_{\sigma j}(t)).
	\label{eq:fragmentation}
\end{equation}
Here, the case of $S_{\sigma}=0$ means that the subsystem $\sigma$ is not depleted, meaning that all particles occupy the same single particle state, i.e. $n_{\sigma 1}=1$.\par
\begin{figure}[t]
	\includegraphics[width=0.8\columnwidth]{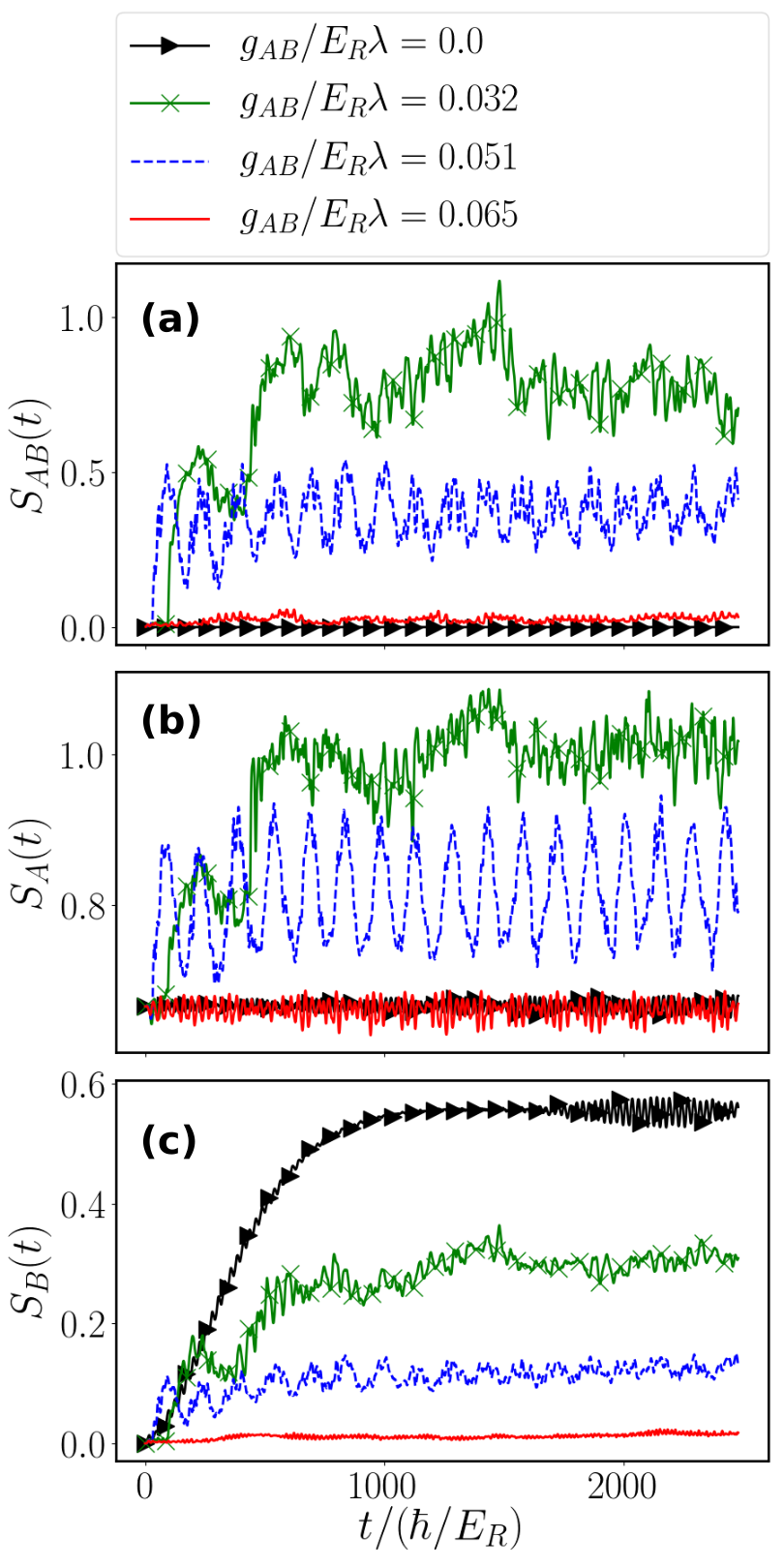}
	\caption{Temporal evolution of (a) the  von Neumann entropy $S_{AB}(t)$, (b) the fragmentation of the impurity species $S_{A}(t)$ and (c) the fragmentation of the the majority species $S_{B}(t)$ for the post-quench interspecies interaction strengths discussed in Figure \ref{fig:densities} corresponding to the four emerging regimes. $t$ is given in units of $\hbar/E_R$.}
	\label{fig:correlations}
\end{figure}
Figure \ref{fig:correlations} shows the temporal evolution of the von Neumann entropy $S_{AB}(t)$ and the fragmentation of the impurity species $S_{A}(t)$ as well as of the majority species $S_{B}(t)$ for the four different regimes of the dynamical response of our system, depending on the post-quench interspecies interaction strength $g_{AB}$.
Again, we start by analyzing the results for the weak quench to $g_{AB}/E_R\lambda=0.065$. Analogously to the one-body density of the impurity species the fragmentation $S_A(t)$ is in general dynamically stable, while showing only marginal fluctuations. The same is true for the von Neumann entropy $S_{AB}(t)$ as well as the fragmentation of the majority species $S_{B}(t)$. Quenching to $g_{AB}/E_R\lambda=0.051$ leads to a strong response of the aforementioned quantities. The initial tunneling of the impurity species into the third well is accompanied by an increase of the fragmentation $S_A(t)$ and $S_B(t)$ as well as the von Neumann entropy $S_{AB}(t)$, whereas the reduction of the impurity species' density in the third well will reduce said quantities. We note that this effect is the strongest for the von Neumann entropy and the fragmentation of species A. The impact on the fragmentation of the majority species is minor and appears to stabilize in the course of time. Initially, the subsystems are disentangled, i.e. $S_{AB}(t=0)=0$ , which manifests itself in the system being described by a single product state $|\Psi_0\rangle\approx|2,2,0,0,0\rangle_W\otimes|\Psi^{3}_B\rangle$.
The moment the impurity species tunnels into the third well it will become entangled with the majority species, while in the case when the impurity species tunnels back the entanglement is reduced. However, in this case the von Neumann entropy does not drop to $S_{AB}=0$, but rather the subsystems remain entangled to some extent. This is due to the fact that a fraction of the impurity species density remains in the third well, thereby increasing the overlap with the majority species, even though it is not clearly visible in the one-body density in Figure \ref{fig:densities} (c). This will become evident in the following discussion of Figure \ref{fig:densities} (e).  \par
For stronger quenches, i.e. $g_{AB}/E_R\lambda=0.032$, the subsystems will get even more entangled, while at the same time the impurity species exhibits stronger correlations. However, these correlations as well as the von Neumann entropy evolve less regularly compared to the previously discussed post-quench interspecies interaction strength. As expected, when quenching to $g_{AB}=0$ and thereby making the majority species transparent to the impurity species the subsystems remain disentangled in the course of time. Since the impurity species remains localized in the first and second well, also the fragmentation $S_A$ only weakly deviates from the initial value with time. In contrast to that, the impact of this quench on the fragmentation of the subsystem B is rather strong, but shall not be the focus of the current work's analysis.  
In summary, we have found four different regimes for the dynamical response of the binary mixture upon quenching the interspecies interaction strength. In the regimes in which the impurity species undergoes a tunneling to the neighbouring wells, we observe strong entanglement between the majority and the impurity species as well as correlations among the impurity atoms. 

\subsection{Microscopic Characterization of the Tunneling Process}
\label{sec:character}
This section is dedicated to an in-depth analysis of the many-body wave function, in particular focusing on the regimes where the impurity species undergoes a tunneling to the neighbouring wells. In a first step, we want to analyze the probability of finding an impurity particle in a specific single particle state. Naturally, for the impurity species we consider generalized Wannier functions, which are obtained by diagonalizing the position operator in the eigenbasis of the single-particle Hamiltonian $H^{(1)}_A=-\frac{\hbar^2}{2m_A}\frac{d^2}{dx^2}+V_0 \sin^2(\frac{\pi k x}{L})$, while being restricted to the lowest band. In a second step, we build number states using the Wannier functions and project the time-dependent full many-body wave function onto these number states, as described in equation \ref{eq:prob_A}. Again, the reader should note that the complete many-body wave function is obtained via ML-MCTDHX and subsequently we analyze this high-dimensional object by projecting onto the corresponding number states.
In order to obtain the probability of finding an impurity particle in a specific Wannier state, we construct the operator
\begin{equation}
	\hat{O}^{(1)}_l=\frac{1}{N_A}\sum^{N_A}_{i} |w^{i}_l\rangle\langle w^{i}_l|,
	\label{eq:w_proj}
\end{equation}
where $|w^{i}_l\rangle\langle w^{i}_l|$ projects the $i-$th particle of the A species onto the $l-$th Wannier state. Evaluating this operator with respect to the complete many-body wave function yields the probability $P_l=\langle\Psi_{\text{MB}}|\hat{O}^{(1)}_l|\Psi_{\text{MB}}\rangle$ of finding an impurity particle in the $l-$th Wannier state. In the following the Wannier states are ordered from left to right, i.e. $|w_1\rangle$ and $|w_2\rangle$ describe the Wannier states which are associated with the initially ($t=0$) populated wells. The Wannier states prove to be a suitable basis set, since in all cases analyzed in the following, we find that $\sum_l P_l\approx99.97\%$.
Figure \ref{fig:densities} (a), (e), show the probability $P_l$ of finding an impurity particle in the $l-$th Wannier state upon quenching $g_{AB}$. For a weak quench, i.e. $g_{AB}/E_R\lambda=0.065$, a single impurity is either to be found in the first or second Wannier state (not shown here), and therefore in the first or second well. For larger quenches, $g_{AB}/E_R\lambda=0.051$, [cf. Figure \ref{fig:densities} (a)] we mainly find oscillations between a single impurity populating the second and third well, which confirm our assumptions regarding the tunneling discussed in section \ref{sec:Tunneling Dynamics}. We additionally find small fluctuations for populating the first Wannier state, which were not directly visible in the respective one-body density of the impurity species.
Finally, quenching to $g_{AB}/E_R\lambda=0.032$ a single impurity will in the course of time populate the fourth Wannier state, i.e. the fourth well, apart from populating the Wannier state one to three. Apparently, the tunneling is of such nature that mainly the second well is depleted, whereas the probability of finding an impurity in the first well only slightly reduces [cf. Figure \ref{fig:densities} (e)]. Using these quantities we gain a first quantitative description of the tunneling behaviour of the impurity species in terms of Wannier states.\par
There remains the question of how many particles actually tunnel to the neighbouring wells. Such an insight can be gained by analyzing $P(|\vec{n}^{A}\rangle)(t)$ (see equation \ref{eq:prob_A}).  We will in particular focus on the two post-quench interspecies interaction strengths, $g_{AB}/E_R\lambda=0.032$ and $g_{AB}/E_R\lambda=0.051$, featuring a tunneling of the impurity species.
\begin{figure}[t]
	\includegraphics[width=0.8\columnwidth]{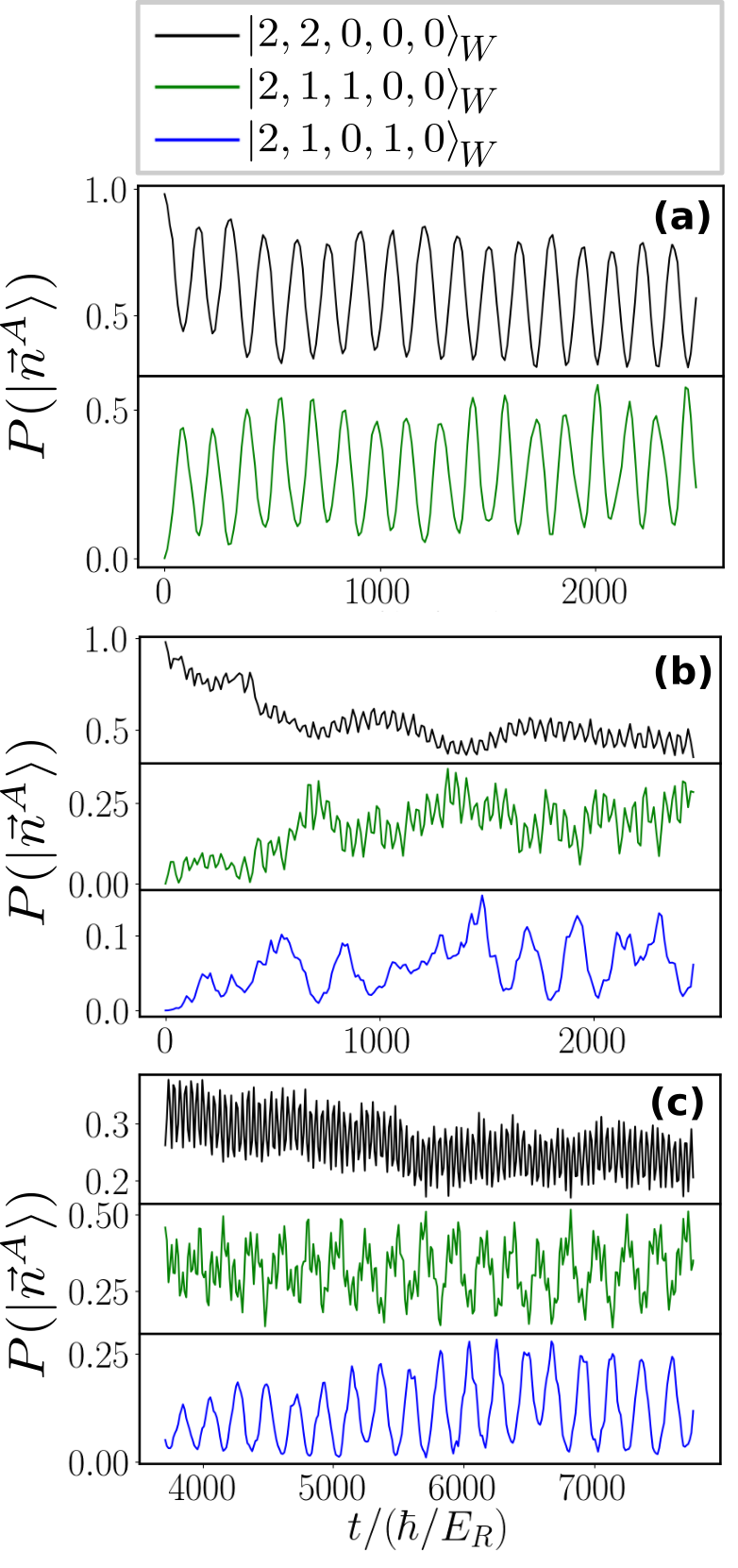}
	\caption{Temporal evolution of the probability $P(|\vec{n}^{A}\rangle)(t)$ of finding the impurity species in the number state $|2,2,0,0,0\rangle_W$ [panel (a), (b), (c) first row], $|2,1,1,0,0\rangle_W$ [panel (a), (b), (c) second row] and $|2,1,0,1,0\rangle_W$ [panel (b), (c) third row], irrespective of the number state configurations of the B species upon quenching to (a) $g_{AB}/E_R\lambda=0.051$ and (b), (c) $g_{AB}/E_R\lambda=0.032$. Panel (c) shows the long time evolution for a quench to $g_{AB}/E_R\lambda=0.032$. The particle number of the respective species is chosen as $N_A=4$ and $N_B=10$. $t$ is given in units of $\hbar/E_R$.}
	\label{fig:number_states_A_evolution}
\end{figure}
In Figure \ref{fig:number_states_A_evolution} we show the probability $P(|\vec{n}^{A}\rangle)(t)$ of finding the impurity species in the number state $|\vec{n}^{A}\rangle$ upon quenching to $g_{AB}/E_R\lambda=0.032$ and $g_{AB}/E_R\lambda=0.051$. We find that the number states with the largest contribution to the complete many-body wave function $|\Psi_{\text{MB}}\rangle$ are given by $|2,2,0,0,0\rangle_W$, $|2,1,1,0,0\rangle_W$ and $|2,1,0,1,0\rangle_W$, whereas for the quench to  $g_{AB}/E_R\lambda=0.051$ predominantly the first two states contribute.
For $t=0$ the impurities accumulate pairwise in adjacent sites and the corresponding many-body wave function is well described by $|\Psi_0\rangle\approx|2,2,0,0,0\rangle_W\otimes|\Psi^{3}_B\rangle$. Directly after the quench the probability of finding the impurities in the configuration $|2,2,0,0,0\rangle_W$ reduces, while the probability of finding the impurities in the configuration $|2,1,1,0,0\rangle_W$ increases. In essence, this means that a single impurity out of the cluster of impurities tunnels to the third well. In that way we have identified the tunneling process, discussed in section \ref{sec:Tunneling Dynamics}, as a single particle process. Focusing now on the stronger quench to $g_{AB}/E_R\lambda=0.032$, the increased density in the fourth well [cf. Figure \ref{fig:densities} (b)] can be again associated with a single particle tunneling into that well. Considering solely the time evolution in Figure \ref{fig:number_states_A_evolution} (b) one might assume that the tunneling to the fourth well (blue line), i.e. the probability of finding the impurity species in the number state $|2,1,0,1,0\rangle_W$ shows a rather irregular behaviour. However, analyzing the long-time evolution of the respective probabilities [Figure \ref{fig:number_states_A_evolution} (c)] one finds a more regular behaviour of the aforementioned probabilities. It appears that the single impurity tunnels between the third and fourth well, while the probability of finding the impurities in the configuration $|2,2,0,0,0\rangle_W$ is strongly reduced and exhibits high-frequency fluctuations.\par
Employing the projection onto number states, we obtained a detailed description of the actual character of the tunneling process in the impurity species. The initial pairwise accumulated impurity cluster breaks up in such a way that a single impurity is transferred to the neighbouring well. Increasing the strength of the quench it is even possible to transfer this single impurity to the fourth well. We note that this effect is due to a finite coupling to the majority species during the dynamics and cannot be achieved when the impurity species is not coupled to the majority species in the course of the dynamics, i.e. $g_{AB}=0$. \par
\begin{figure}[t]
	\includegraphics[width=0.9\columnwidth]{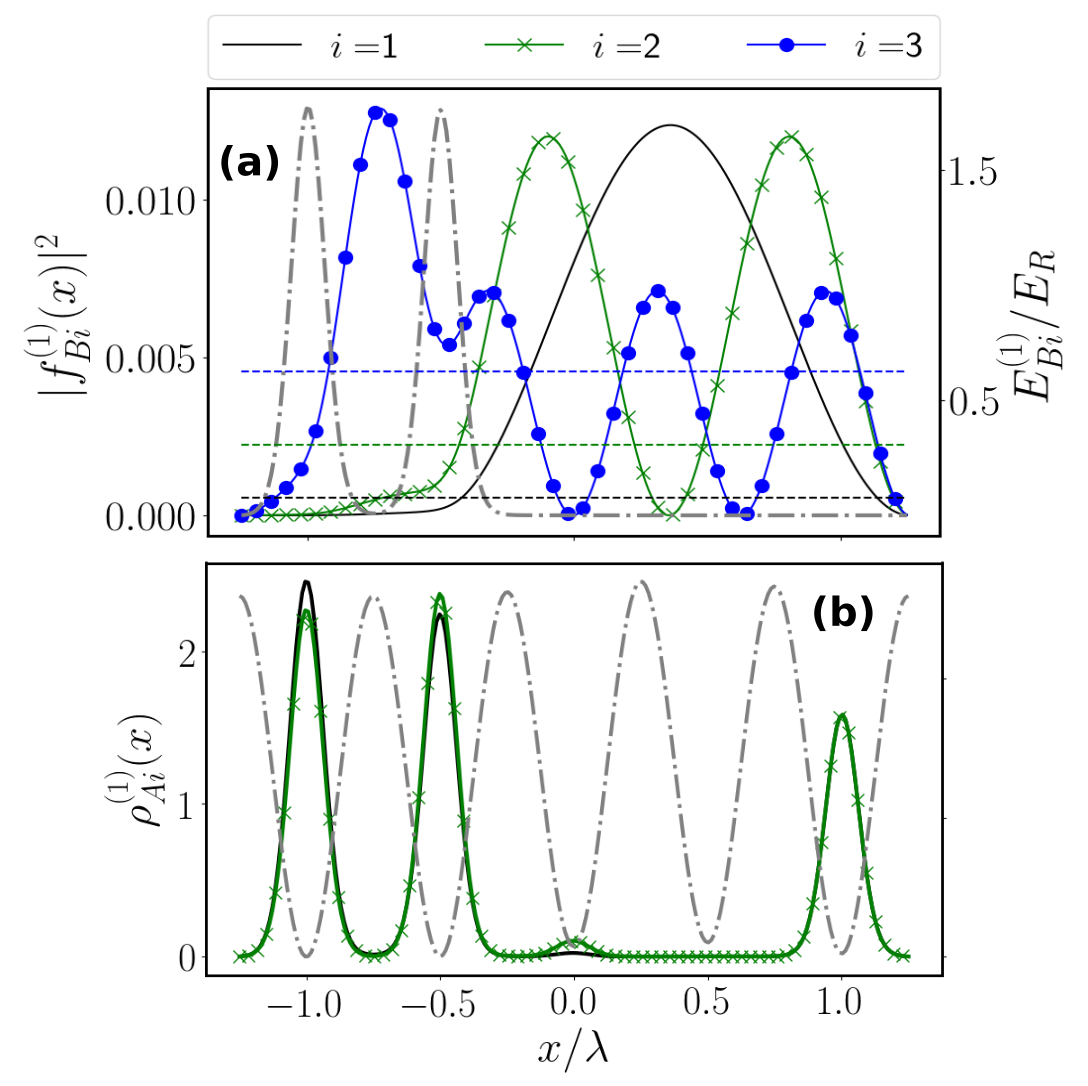}
	\caption{(a) Spatial resolution of the eigenvectors $f^{(1)}_{B i}(x)$ of the effective single-particle Hamiltonian $\hat{H}^{(1)}_B$ (equation \ref{eq:eff_h_B}) and their respective eigenenergies $E^{(1)}_{B i}$ (dashed lines). (b) One body-density of the many-body ground state ($i=1$) and first excited state ($i=2$) w.r.t the Hamiltonian $\hat{H}_{A,eff.}$. The dash-dotted grey line in (a) refers to the potential $g_{AB}N_A\rho^{(1)}_A(x)$ (with pre-quench $g_{AB}/E_R\lambda=0.142$) and in (b) to $g_{AB}N_B\rho^{(1)}_B(x)+V_0 \sin^{2}\Big(\frac{\pi k x_i}{L}\Big)$ (with post-quench $g_{AB}/E_R\lambda=0.051$) in the respective effective Hamiltonian with $V_0/E_R=13$. The particle number of the respective species is chosen as $N_A=4$ and $N_B=10$. $x$ is given in units of $\lambda$ and $E^{(1)}_{Bi}$ in units of $E_R$.}
	\label{fig:eigvec_B_eff_lattice}
\end{figure}

Since the majority species plays an important role for the tunneling process it is of imminent interest to understand in which way this species is excited due to the quench. For this purpose, we aim at projecting the complete many-body wave function onto number states $|\vec{n}^{A}\rangle\otimes|\vec{n}^{B}\rangle$, but rather determine the probability of being in the number state $|\vec{n}^{B}\rangle$ irrespective of the number state configurations of the impurity species. Before this can be done we need to find an appropriate single particle basis set upon which to build the number states  $|\vec{n}^{B}\rangle$, which is defined as 
\begin{equation}
	P(|\vec{n}^{B}\rangle)= \sum_{i} |\langle \vec{n}^{B} |\otimes\langle \vec{n}^{A}_i|\Psi_{\text{MB}}\rangle|^{2}.
\end{equation}
Instead of using the single particle functions associated with the geometry, i.e the eigenstates of a single particle in a box (hard wall boundary conditions), we rather want to employ a basis which is more tailored to the many-body problem and exhibits an impact of the interspecies interaction. For this reason we determine the effective single particle Hamiltonian for the B species, assuming the product ansatz $|\Psi_0\rangle\approx|2,2,0,0,0\rangle_W\otimes|\Psi^{3}_B\rangle$ of our ground state for $t=0$. Integrating out the impurity species we arrive at the following effective non-interacting single-particle Hamiltonian for the B species
\begin{equation}
	\hat{H}^{(1)}_B=-\frac{\hbar^2}{2m_B}\frac{d^2}{dx^2}+g_{AB}N_A\rho^{(1)}_A(x,t=0).
	\label{eq:eff_h_B}
\end{equation}
Through diagonalization of $\hat{H}^{(1)}_B$ in a discrete variable representation basis \cite{Light1985} we obtain an eigenbasis for the majority species which takes the density modulation due to the presence of the impurity species into account. In Figure \ref{fig:eigvec_B_eff_lattice} (a) we show the corresponding eigenstates $f^{(1)}_{Bi}(x)$ together with their eigenenergies $E^{(1)}_{Bi}$.
\begin{figure}[t]
	\includegraphics[width=0.8\columnwidth]{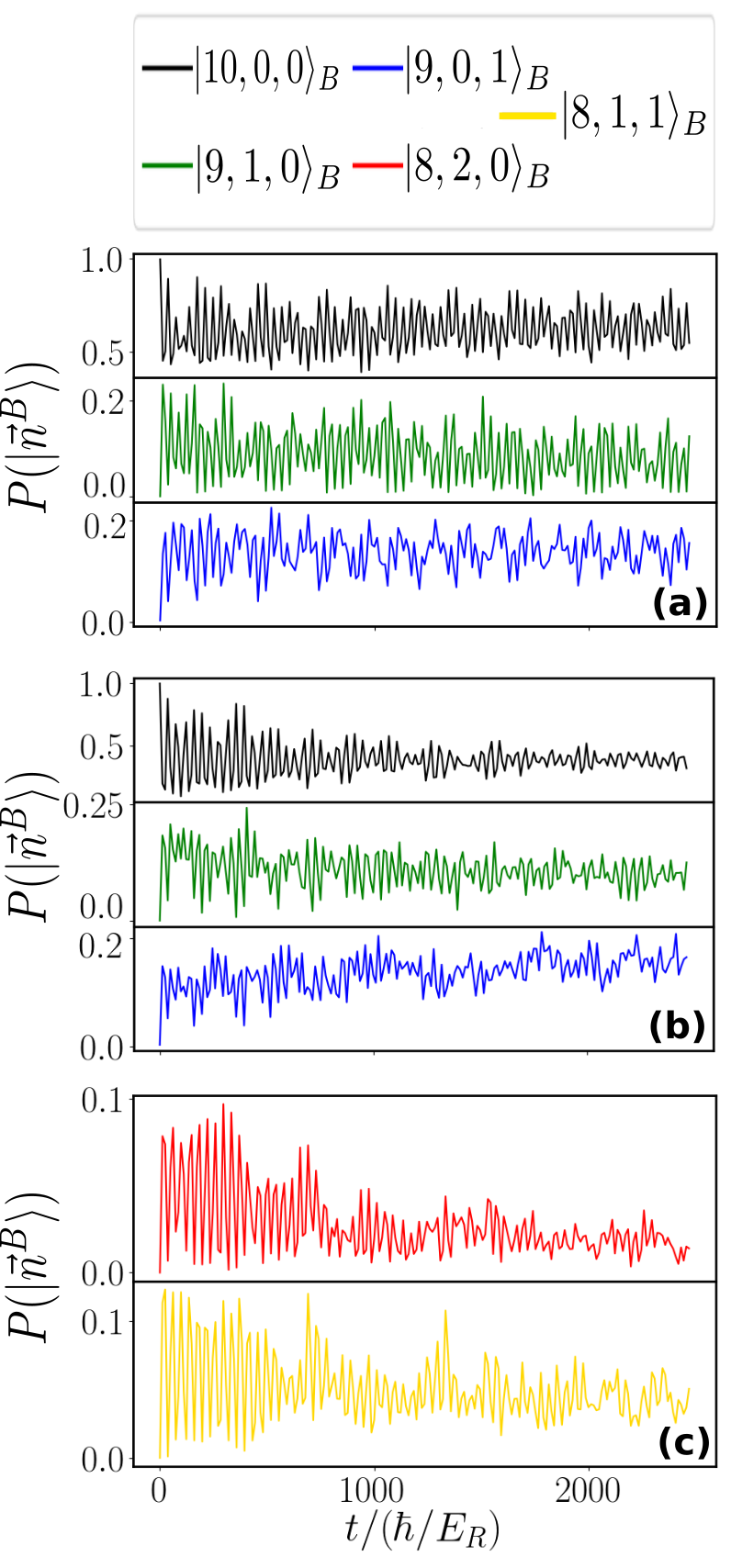}
	\caption{Temporal evolution of the probability $P(|\vec{n}^{B}\rangle)(t)$ of finding the majority species in the number state $|10,0,0\rangle_B$ [panel (a), (b) first row], $|9,1,0\rangle_B$ [panel (a), (b) second row], $|9,0,1\rangle_B$ [panel (a), (b) third row], $|8,2,0\rangle_B$ [panel (c) first row] and $|8,1,1\rangle_B$ [panel (c) second row], irrespective of the number state configurations of the A species upon quenching to (a) $g_{AB}/E_R\lambda=0.051$ and (b), (c) $g_{AB}/E_R\lambda=0.032$. Panel (c) shows the probability for two-particle excitations for the latter quench. The particle number of the respective species is chosen as $N_A=4$ and $N_B=10$. $t$ is given in units of $\hbar/E_R$.}
	\label{fig:number_state_evolution_B}
\end{figure}
The single-particle ground state wave function $f^{(1)}_{B1}(x)$ is localized such that it strongly avoids the effective potential imposed by the impurity species. Apart from a slightly larger overlap with the potential $g_{AB}N_A\rho^{(1)}_A(x)$, the first excited state $f^{(1)}_{B2}(x)$ also avoids the latter, but additionally it exhibits a single node. The next energetically higher state $f^{(1)}_{B3}(x)$ is distributed between the peaks of the effective potential, as well as in the region where the potential is zero, having two nodes in total. \par
Using these single-particle functions we build the corresponding number states $|\vec{n}^{B}\rangle$ and determine analogously to the previous analysis $P(|\vec{n}^{B}\rangle)(t)$.
Again we only focus on the most populated number states in the following. We note that excitations into higher excited states are negligible and it is sufficient to consider only three single-particle states for the construction of the number states. In Figure \ref{fig:number_state_evolution_B} we observe that the temporal evolution of the probability $P(|\vec{n}^{B}\rangle)(t)$ of finding the majority species in a specific number state is dominated by single particle excitations, i.e. $|9,1,0\rangle_B$ and $|9,0,1\rangle_B$. In this sense, the tunneling of the impurity into the neighbouring sites is correlated with single-particle excitations in the majority species. Compared to the rather regular behaviour of excitations in the impurity species, the excitations in the majority species fluctuate strongly which is related to the high-frequency oscillations of the corresponding one-body density. The initial species wave function of the B species is given by $|\Psi^{3}_B\rangle\approx|10,0,0\rangle_B$. Following the quench, the population of the number state $|10,0,0\rangle_B$ is reduced dramatically in the course of time. Additionally, we observe two-particle excitations which contribute strongly shortly after the quench, but reduce in the course of time to a few percent [cf. Figure \ref{fig:number_state_evolution_B} (c)]. A similar observation is made for a quench to $g_{AB}/E_R\lambda=0.051$, but with a lowered contribution of two-particle excitations. 
Concluding, we have found that the tunneling of a single impurity to the neighbouring wells of the lattice is accompanied by single-particle excitations in the majority species with respect to the eigenstates of an effective Hamiltonian. \par
Finally, using the obtained knowledge about the microscopic processes we give an explanation of the origin of the frequency content of the one-body densities $\rho^{(1)}_\sigma(x,t)$ in Figure \ref{fig:densities}, and in particular unravel the nature of the different time scales of the corresponding oscillations. In the following we focus on the quench to $g_{AB}/E_R\lambda=0.051$, where we observe tunneling of a single impurity solely to the third well \ref{fig:densities} (c). The employed effective Hamiltonians are constructed w.r.t the post-quench interspecies interaction strength (i.e. $g_{AB}/E_R\lambda=0.051$) which appears in $g_{AB}N_\sigma\rho^{(1)}_\sigma(x)$, $\sigma \in\{A,B\}$. In $\rho^{(1)}_B(x,t)$ [\ref{fig:densities} (g)] we observed high frequency oscillations upon quenching to $g_{AB}/E_R\lambda=0.051$. We can identify two dominant fundamental frequencies with single-particle excitations from the state $f^{(1)}_{B1}(x)$ to $f^{(1)}_{B2}(x)$ with $(E^{(1)}_{B2}-E^{(1)}_{B1})/E_R\approx0.18$ and from $f^{(1)}_{B1}(x)$ to $f^{(1)}_{B3}(x)$ with $(E^{(1)}_{B3}-E^{(1)}_{B1})/E_R\approx0.37$. This single-particle picture is justified by the fact that the particles of the majority species are interacting weakly with each other. However, the fundamental frequency of $\rho^{(1)}_A(x,t)$ w.r.t the tunneling into the thir well cannot be identified in a single-particle picture since the impurities interact rather strongly via an intraspecies interaction of strength $g_{AA}/E_R \lambda=0.067$. For this reason, we construct a many-body Hamiltonian for the impurity species by integrating out the majority species analogously to equation \ref{eq:eff_h_B}, yielding
\begin{equation}
	\begin{split}
		\hat{H}_{A,eff.}&=\sum_{i=1}^{N_A}\Big[ -\frac{\hbar^2}{2m_A}\frac{d^2}{dx_i^2}+g_{AB}N_B\rho^{(1)}_B(x_i,t=0)\\
		&+V_0 \sin^{2}\Big(\frac{\pi k x_i}{L}\Big)\Big]+g_{AA}\sum_{i<j}\delta(x_i-x_j).
	\end{split}
\end{equation}
In Figure \ref{fig:eigvec_B_eff_lattice} (b) we present the one-body density of the ground state ($i=1$) and first excited state ($i=2$) of the many-body Hamiltonian $\hat{H}_{A,eff.}$. We find that the fundamental frequency of $\rho^{(1)}_A(x,t)$ corresponds to a transition from the ground state to the first excited state with $(E_{A2}-E_{A1})/E_R\approx0.043$. In the one-body density of the excited state this manifests itself in a higher density in the third and fifth well of the effective potential, thereby indicating the observed tunneling of the impurity species.

\subsection{Increasing the Number of Majority Species Particles}
\begin{figure}[h!]
	\includegraphics[width=0.8\columnwidth]{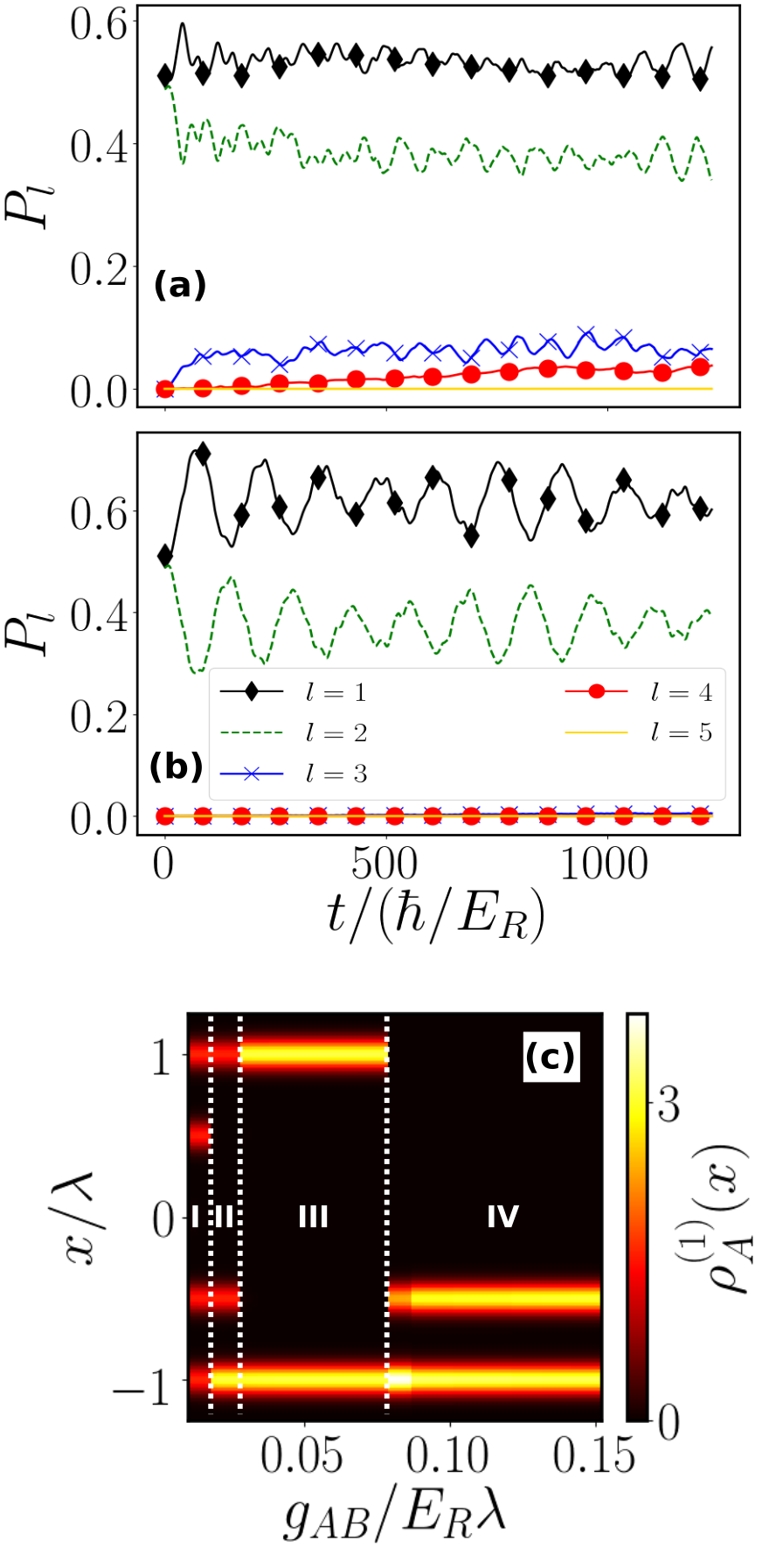}
	\caption{Temporal evolution of the probability $P_l$ of finding an impurity particle in the $l-$th Wannier state upon quenching to (a) $g_{AB}/E_R\lambda=0.03$ and (b) $g_{AB}/E_R\lambda=0.051$. (c) One-body density $\rho^{(1)}_A(x)$ of the species A as a function of the interspecies interaction strength for a fixed lattice depth of $V_0/E_R=13$ (see Table \ref{table_statesNB30}). The particle number of the respective species is chosen as $N_A=4$ and $N_B=30$. $x$ is given in units of $\lambda$, $t$ in units of $\hbar/E_R$ and $g_{AB}$ in units of $E_R \lambda$.}
	\label{fig:single_wannier_NB30}
\end{figure}

\begin{table*}[t]
	\captionof{table}{Degenerate subspaces of the ground state referring to the regions in Figure \ref{fig:single_wannier_NB30} (c) for $V_0/E_R=13$ and $N_B=30$ particles in the majority species.}
	\label{table_statesNB30}
	\begin{tabular}{cccc}
		\hline\hline
		Regime: $I$  & $II$ &  $III$ & $IV$\\\\ \hline
		$|1,1,0,1,1\rangle_W\otimes|\Psi^1_B\rangle$  $\;$& $|2,1,0,0,1\rangle_W\otimes|\Psi^{2}_B\rangle$  $\;$& $|2,0,0,0,2\rangle_W\otimes|\Psi^{3}_B\rangle$  $\;$& $|2,2,0,0,0\rangle_W\otimes|\Psi^{4}_B\rangle $ \\\\
		&  $|1,0,0,2,1\rangle_W\otimes|\bar{\Psi}^2_B\rangle$  $\;$&  $\;$& $|0,0,0,2,2\rangle_W\otimes|\bar{\Psi}^4_B\rangle$ \\
		\hline\hline
	\end{tabular}
\end{table*}
Let us now find out whether the observed tunneling processes for $N_B=10$ particles in the majority species persist if the number of particles in the majority species is increased. For this purpose, we set the latter to $N_B=30$ and perform a similar analysis as compared to the previous sections. Before we turn to the dynamical response of the binary mixture let us first investigate the ground state wave function for varying interspecies interaction strength. Instead of analyzing the explicit form of the wave function by projecting onto the corresponding number states as done in section \ref{sec:Quench Protocol} we investigate the one-body density of the impurity species. In Figure \ref{fig:single_wannier_NB30} (c) we show the one-body density of the impurity species for varying interspecies interaction strengths $g_{AB}$. We find four different regimes for the distribution of the impurities in the lattice, whereas for $N_B=10$ the ground state wave function was classified by three regimes. The additional regime is characterized by impurities accumulated pairwise in the outermost wells, while the majority species occupies the central region (cf. \cite{keiler2}). The corresponding regimes, taking into account the degenerate subspaces of the wave function, are summarized in Table \ref{table_statesNB30}.
We now prepare our system with the same parameters as in the previous sections, i.e. we choose a lattice depth of $V_0/E_R=13$ and an interspecies interaction strength of $g_{AB}/E_R\lambda=0.14$. This leads to an initial wave function $|\Psi_0\rangle\approx|2,2,0,0,0\rangle_W\otimes|\Psi^{4}_B\rangle$ of the same form as in section \ref{sec:Tunneling Dynamics}. Again by quenching the interspecies interaction strength to lower values we initiate the tunneling dynamics of the impurity species. 
In order to quantify the tunneling of the impurity species we analyze the probability $P_l=\langle\Psi_{\text{MB}}|\hat{O}^{(1)}_l|\Psi_{\text{MB}}\rangle$ of finding an impurity particle in the $l-$th Wannier state. In Figure \ref{fig:single_wannier_NB30} we show the temporal evolution of $P_l$ upon quenching to (a) $g_{AB}/E_R\lambda=0.03$ and (b) $g_{AB}/E_R\lambda=0.051$. For a quench to $g_{AB}/E_R\lambda=0.03$ we observe a tunneling of the impurity species to the third and fourth site, while being slightly reduced in the intensity [\ref{fig:single_wannier_NB30} (a)] compared to the case of $N_B=10$ particles [cf. Figure \ref{fig:densities} (b)]. However, for a quench to $g_{AB}/E_R\lambda=0.051$ we solely observe a tunneling of the impurity species between the initially populated wells [\ref{fig:single_wannier_NB30} (b)].  A reduction of the probability of finding a single impurity in the first well is accompanied by an increase of that probability in the second well and vice versa. Compared to the previous study with $N_B=10$ particles in the majority species we do not find scenarios with a tunneling of the impurities which is restricted to the the third well [cf. Figure \ref{fig:densities} (c)].  Apparently, the increased repulsive interspecies interaction with the majority species due to the increased particle number leads to the fact that the impurity species will be forced to populate the outermost well stronger compared to the case of $N_B=10$ particles. Nevertheless, quenching to $g_{AB}/E_R\lambda=0$ still results in a localization of the impurities in the initially populated wells in the course of time (not shown here). \par 
In conclusion, we have found that increasing the number of particles in the majority species leads to an altered crossover diagram where an additional regime appears. The dynamical response of the binary mixture changes due to the increased number of majority species particles. However, it is still possible to transfer the impurity species through the lattice by quenching the interspecies interaction strength. Again, this effect is due to the presence of the majority species, i.e. a finite coupling to the latter, and does not occur when the majority species is transparent to the impurities, i.e. $g_{AB}=0$. Quenching to  $g_{AB}=0$ barely alters the initial distribution of the impurities, such that they remain localized throughout the dynamics.

\section{Conclusions and Outlook}
\label{sec:Conclusion}
We have demonstrated that it is possible to transfer a single impurity out of initially clustered impurities through a lattice by coupling this species to a majority species. Starting from four  impurities which accumulate pairwise in adjacent lattice sites we have quenched the interspecies interaction strength to a lower value. Utilizing this quench protocol we cross the boundaries of the crossover diagram of the ground state. The different regimes show different impurity distributions as a function of the interspecies interaction strength. For sufficiently small post-quench interspecies interaction strengths we observe the tunneling of a single impurity out of the cluster, whereas a quench to $g_{AB}=0$ leads to a localization of the impurities in the initially populated wells. The tunneling process of the impurity species is accompanied by strong entanglement of the subsystems as well as strong correlations among the impurities. The effect on the majority species manifests itself predominantly in single particle excitations with respect to the eigenstates of an effective Hamiltonian which accounts for the initial one-body density distribution of the impurity species. In contrast to this we find that a single impurity does not lead to a controlled tunneling in the presence of a majority species. 
Furthermore, we have investigated the robustness of the tunneling process by increasing the number of particles in the majority species and revealed that it is still possible to transfer the impurity species through the lattice by quenching the interspecies interaction strength in case of a larger majority species. This controlled ejection of a single impurity may indeed be of interest for future applications in atomtronics \cite{atomtronics}. In this context, it might pave new pathways, since we explicitly take advantage of the coupling to an environment, whereas usually such systems would lose their coherence when coupled to an environment due to dissipation. \par There are several possible directions of future investigations. For example, it would be of immediate interest to allow for a spin degree of freedom in the impurity species, such that the initial distribution is not only characterized by a spatial distribution of the impurities, but also by a spin distribution. In this spirit, a quench of the interspecies interaction strength might lead to a spatial transport of the impurities as well as a spin transport.

\begin{acknowledgements}
	K.K. thanks M. Pyzh for many insightful discussions. P. S. gratefully acknowledges funding by the Deutsche Forschungsgemeinschaft in the framework of the SFB 925 "Light induced dynamics and control of correlated quantum systems" and support by the excellence cluster "The Hamburg Centre for Ultrafast Imaging-Structure, Dynamics and Control of Matter at the Atomic Scale" of the Deutsche Forschungsgemeinschaft. K. K. gratefully acknowledges a scholarship of the Studienstiftung des deutschen Volkes.
\end{acknowledgements}

\appendix*
\section{Convergence Analysis Within ML-MCTDHX}
\label{sec:Appendix}
\begin{figure}[t]
	\includegraphics[width=0.8\columnwidth]{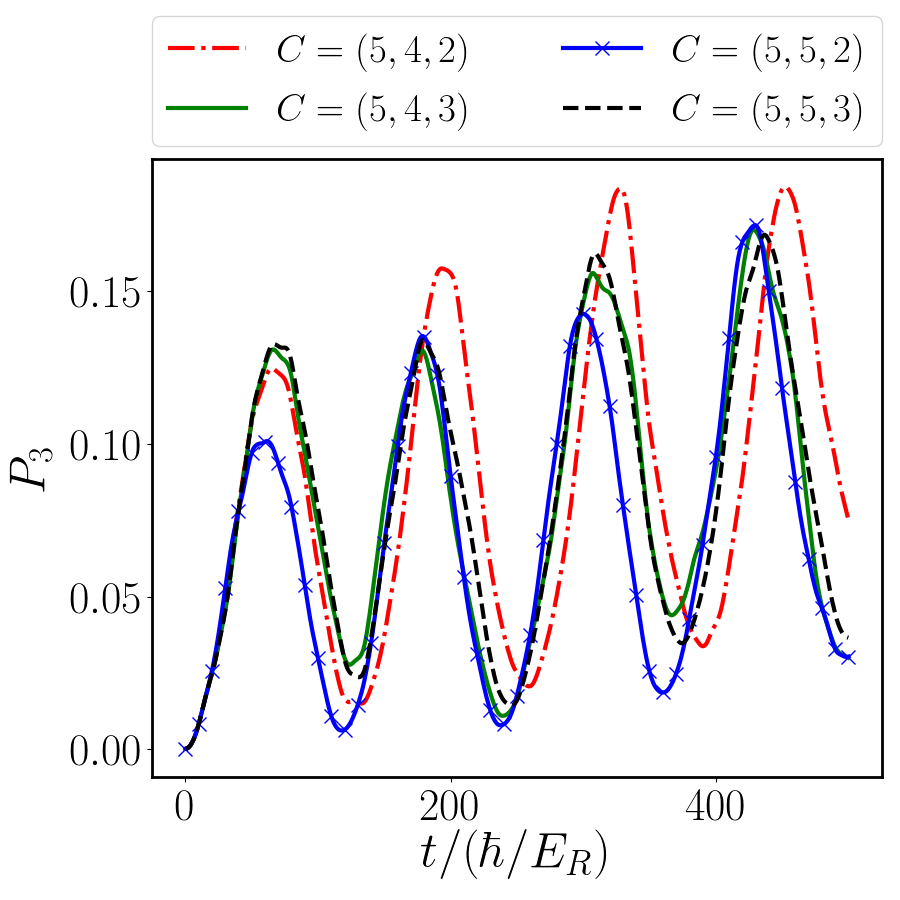}
	\caption{Temporal evolution of the probability $P_3$ of finding an impurity particle in the third Wannier state upon quenching to $g_{AB}/E_R\lambda=0.051$ for $N_B=10$ particles in the majority species. $t$ is given in units of $\hbar/E_R$}
	\label{fig:convergence}
\end{figure} 
In the following, we briefly discuss the convergence of our results using a specific example case and the necessity of a computational approach which is able to take beyond mean-field effects into account for the description of the dynamical response of our binary mixture. The degree of truncation of the underlying Hilbert space is given by the orbital configuration $C=(M,d_A,d_B)$. Here, $M$ refers to the number of species functions in the Schmidt decomposition (cf. equation \ref{eq:schmidt}), while $d_\sigma$ with $\sigma \in\{A,B\}$ denote the number of single-particle functions spanning the time-dependent number states $ |\vec{n}^\sigma; t\rangle$ (cf. equation \ref{eq:ml_ns}). For $M=d_A=d_B=1$ we obtain the solution of the Gross-Pitaevskii mean-field approximation. Increasing the number of species functions as well as single-particle functions we are able to recover the solution of the many-body quantum system with an increasing degree of accuracy. However, choosing too many species and single-particle functions is computationally prohibitive. Nevertheless, numerical solutions which incorporate the relevant correlations and go beyond mean-field approximations can be obtained using ML-MCTDHX. In order to determine the effect of the truncation, we investigate as a representative example the probability $P_3$ of finding the impurity species in the third well (cf. equation \ref{eq:w_proj}) upon varying the orbital configuration $C$. In Figure \ref{fig:convergence} we show this for example for a quench to $g_{AB}/E_R\lambda=0.051$ and $N_B=10$ particles in the majority species. 
We observe that it is possible to achieve convergence by systematically increasing the number of species functions $M$ and single-particle functions $d_\sigma$. Therefore, the orbital configuration $C=(5,5,3)$ has been employed for all many-body calculations in the main text, yielding sufficiently converged results of our observables.


\begin{thebibliography}{80}
\bibitem{footnote1}
	In principle it is necessary to consider also Wannier states of higher bands, but it turns out that the many-body ground state wave function is approximately well described by a superposition of tensor products of number states $|\vec{n}^{A}\rangle\otimes|\vec{n}^{B}\rangle$ with the Fock space of the A species restricted to the lowest band. Number states spanned by Wannier states of higher bands do not contribute.
\end{thebibliography}
\end{document}